\documentclass[]{spie}  

\usepackage{amsmath,amsfonts,amssymb}
\usepackage{comment}
\usepackage{graphicx}
\graphicspath{ {./Figures/} }
\usepackage[colorlinks=true, allcolors=blue]{hyperref}

\newcommand{\vk}{von-K\'{a}rm\'{a}n}

\title{Gaussian phase autocorrelation as an accurate compensator for FFT-based atmospheric phase screen simulations}

\author[a]{Sorabh Chhabra}
\author[a,e]{Jyotirmay Paul}
\author[a, c, d]{A. N. Ramaprakash}
\author[b]{Avinash Surendran}
\affil[a]{Inter-University Center for Astronomy And Astrophysics, Pune 411007, India}
\affil[b]{W. M. Keck Observatory,Kamuela, Hawaii 96743, USA}
\affil[c]{Cahill Center for Astronomy and Astrophysics, California Institute of Technology, Pasadena, CA 91125, USA}
\affil[d]{Institute of Astronomy, Foundation for Research and Technology-Hellas, Voutes, 71110 Heraklion, Greece}
\affil[e]{University of Liège, Space Sciences Technologies \& Astrophysics Research Institute, Liège,4000, Belgium}

\authorinfo{Further author information: (Send correspondence to Sorabh Chhabra) E-mail: sorabh@iucaa.in/sorabh.chhabra@gmail.com}

\begin{document} 
\maketitle

\begin{abstract}
Accurately simulating the atmospheric turbulence
behaviour is always challenging. The well-known FFT based method falls
short in correctly predicting both the low and high frequency
behaviours. Sub-harmonic compensation aids in low-frequency correction
but does not solve the problem for all screen size to outer scale
parameter ratios (G/$L_0$). FFT-based simulation gives accurate result only
for relatively large screen size to outer scale parameter ratio (G/$L_0$).
In this work, we have introduced a Gaussian phase autocorrelation matrix
to compensate for any sort of residual errors after applying for a modified
subharmonics compensation. With this, we have solved problems such as
under sampling at the high-frequency range, unequal sampling/weights for
subharmonics addition at low-frequency range and the patch normalization
factor. Our approach reduces the maximum error in phase structure-function in
the simulation with respect to theoretical prediction to within 1.8\%, G/$L_0$ = 1/1000.
\end{abstract}




\section{Introduction}
\label{sec:intro}  

\begin{figure}[h!]
    \centerline{\includegraphics[width=0.75\textwidth]{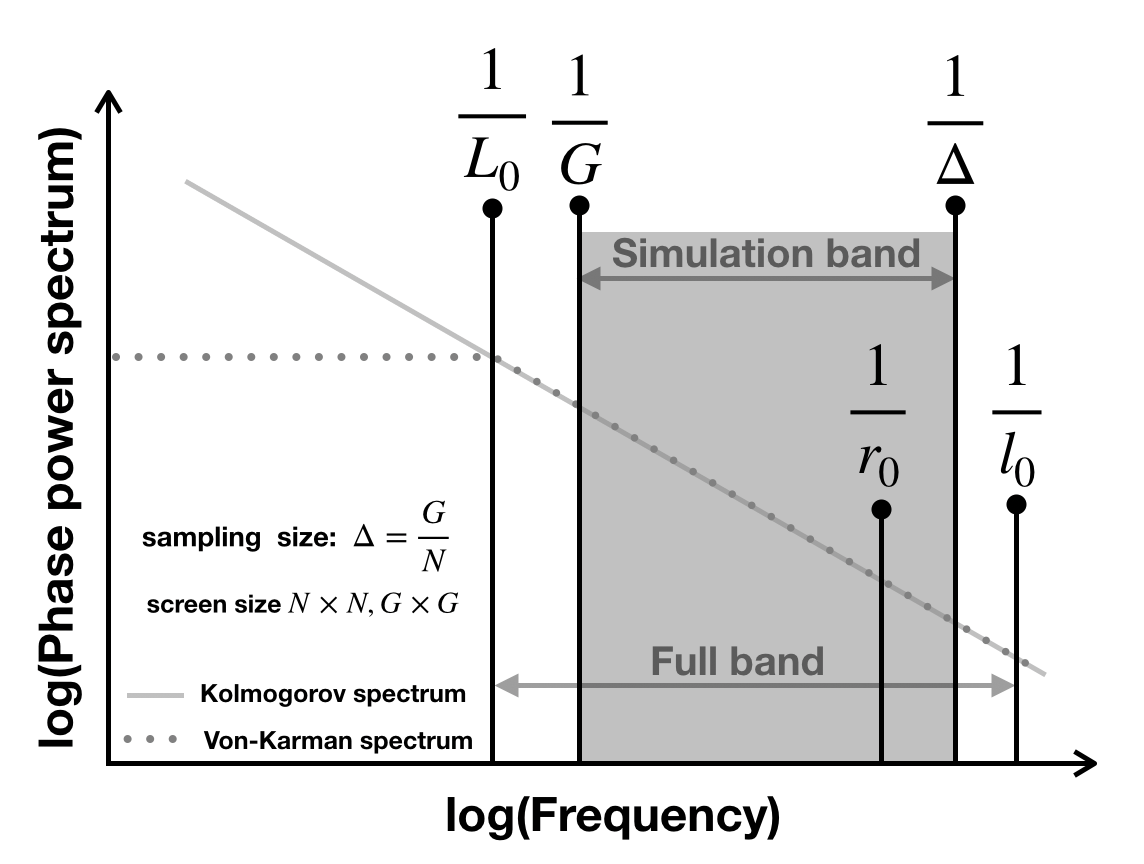}}
    \caption{\label{fig:power}Comparison between Simulation band and Full band}
\end{figure}

For a variety of purposes such as the design and development of adaptive optics systems, speckle imaging techniques, atmospheric propagation studies etc., it is essential to simulate a good atmospheric phase screen model.  Methods based on Zernike polynomial expansions\cite{Roddier90}, FFT-based methods \cite{Herman90,Johansson94,Sedmak98,McGlamery76,Sedmak04,Xiang14,Xiang12}, Optimization method \cite{Zhang19} etc. have been in use for this purpose.The Zernike polynomial method, which is widely in use, has a limitation due to the maximum number of coefficients needed for accurate compensation. The optimization method which compensates accurately for low frequency part of the spectrum by using unequal sampling and unequal weight in low frequency region, but does not cover high frequency deficiencies.  Among these, FFT-based methods are computer memory size friendly and widely accepted. But, FFT operators assume uniform sampling for the non-uniformly distributed phase power spectrum which leads to undersampling in the low and high frequency region. It has limitations in true recreation of the phase power spectrum. To compensate for low-frequency components, Johansson and Gravel \cite{Johansson94} suggested the modified subharmonics equation (originally given by Lane et al. \cite{Lane92}). Giorgio Sedmek \cite{Sedmak04} later proposed a performance analysis of this method by actually calculating the phase structure function from the simulated screen after compensating for high-frequency component too. Results from his analysis show that FFT-based simulations are accurate for large screen size $G$ to Outer scale parameter $L_0$. For a screen size of $G=200$m and outer scale of $L_0 = 25$m, the maximum relative error in the system approaches 1\%. From Fig ~\ref{fig:power} we can see that\cite{sedmak_private} the simulation band $\big(\frac{1}{G}-\frac{1}{\Delta}\big)$ is actually  smaller than full band $\big(\frac{1}{L_0}-\frac{1}{l_0}\big)$. The larger the simulation band to full band ratio, the more accurate the simulated results will be.

Our simulations also demonstrate that the errors from low-frequency components start shooting up once we move to smaller $G/L_0$ ratios, even after compensating with modified subharmonics.
For simulations of imaging with small apertures relative to the outer scale, we need a screen of small size, but cutting out small screens from a larger screen is not the right solution to this problem. \\
In this paper, we present a method to deal with small $G/L_0$ phase screen simulation using the FFT-based method, inspired by Jingsong Xiang's \cite{Xiang14} work on phase screen simulation. 

Section~\ref{sec:autocorr} covers obtaining Phase Autocorrelation Matrix using Phase power spectrum , 
Section~\ref{sec:comp} covers compensation for residual error in phase autocorrelation matrix,
Section~\ref{sec:phase_screen} covers on phase screen simulation from the autocorrelation matrix,
Section~\ref{sec:val} covers validation via Phase structure function calculation,
Section~\ref{sec:results} covers result analysis and
Section~\ref{sec:conclusion} covers conclusion part.

\section{Obtaining Phase Autocorrelation Matrix using Phase Power Spectrum}\label{sec:autocorr}

The 2D phase structure-function $D_{\phi}(m,n)$ and phase autocorrelation matrix $B_{\phi}(m,n)$ are related as follows:

\begin{equation}\label{eqn:sf_ac}
    D_{\phi}(m,n) = 2(B_{\phi}(0,0)-B_{\phi}(m,n)),
\end{equation}
where $B_{\phi}(m,n)$ is the phase autocorrelation matrix and $(m,n)$ are the coordinates along x and y-axis. 
The 2D phase autocorrelation matrices for the FFT-Based phase screen and the modified subharmonic method by Johansson and Gavel \cite{Johansson94} are represented as follows.

\begin{equation} \label{eqn:ac_FFT}
B_{\phi}^{FFT}(m, n)  =\sum_{m^{\prime}=-N_{x} / 2}^{N_{x} / 2-1} \sum_{n^{\prime}=-N_{y} / 2}^{N_{y} / 2-1} f^{2}_{FFT}\left(m^{\prime}, n^{\prime}\right) e^{i 2 \pi\left(\frac{m^{\prime} m}{N_{x}}+\frac{n^{\prime} n}{N_{y}}\right)} 
\end{equation}

\begin{equation} \label{eqn:ac_SH}
B_{\phi}^{SUB}(m, n)=\sum_{p=1}^{N_{p}} \sum_{m^{\prime}=-3}^{2} \sum_{n^{\prime}=-3}^{2} f_{SUB}^{2}\left(m^{\prime}, n^{\prime}\right) e^{i 2 \pi 3^{-p}\left(\frac{\left(m^{\prime}+0.5\right) m}{N_{x}}+\frac{\left(n^{\prime}+0.5\right) n}{N_{y}}\right)}, 
\end{equation}
where $f^{2}_{FFT}\left(m^{\prime}, n^{\prime}\right)$ and $f_{SUB}^{2}\left(m^{\prime}, n^{\prime}\right)$ are the  \vk\ power spectrum and subharmonic power spectrum of Johansson and Gavel \cite{Johansson94}. $(N_x,N_y)$ are sample points, p is the $p^{th}$ subharmonic and $N_p$ is the total number of subharmonics. During subharmoic addition, we are not worried about the leakage of energy from subharmonics to harmonics. Thus, patch normalization factor, to compensate for this leakage, is set to 1. The leakage compensation part is dealt with in section ~\ref{sec:comp}.

The 2D phase autocorrelation matrix after compensating with subharmonics is represented as
    
\begin{equation}\label{eqn:net_ac}
    B_{\phi}(m,n) = B^{FFT}_{\phi}(m,n) +B^{SUB}_{\phi}(m,n).
\end{equation}
We can determine the phase structure function of the simulated screen by substituting Eqn. \ref{eqn:net_ac} in Eqn. \ref{eqn:sf_ac}. 

\section{Compensation for residual error in phase autocorrelation matrix}\label{sec:comp}
It is clear from the work of Zhang et. al. \cite{Zhang19} on optimum frequency sampling that unequal sampling and unequal weights are the most optimized solution to compensate for the under-sampling problem in the low-frequency region. The subharmonic method follows one particular fashion of sampling ($3^{-p}$) as is seen from Eqn. \ref{eqn:ac_SH}. With the use of Eqn. \ref{eqn:sf_ac} to Eqn. \ref{eqn:net_ac}, first $D_{\phi}(m,n)$ is calculated with the assumption that $B_{\phi}^{FFT}(0,0)$ and $B_{\phi}^{SUB}(0,0)$ are zero because we are not concerned about the piston component for now. The remaining error in $D_{\phi}(m,n)$ with respect to $D_{theory}$ can be calculated as follows:

\begin{equation}\label{eqn:error_D}
        D_{error}(m,n) = D_{theory}(m,n) - D_{\phi}(m,n),  
    \end{equation}
where $D_{theory}(m,n)$ is the theoretical phase structure matrix taken from Johansson and Gavel\cite{Johansson94}.

The error matrix cannot be just added to the $D_{\phi}$ matrix.The reason is the following: Any curve that is represented by a polynomial equation will have a higher order of moments. If we take the Fourier transform of this polynomial equation, the resultant curve will be completely different with a different order of moments, just like Gibbs phenomena. This introduces unwanted error in the final result. During subharmoic addition, we are not worried about the leakage of energy from subharmonics to harmonics. Thus, patch normalization factor, to compensate for this leakage, is set to 1 and this leakage compensation part has been dealt in section ~\ref{sec:comp}. We need a smoothening operator like a Gaussian function that can compensate for the remaining error. Using MATLAB, we find the best fit of $D_{error}(m,n)$ using cftool to obtain coefficients of the required Gaussian function (with 95\% confidence bounds)\footnote{With cftool, first fitting was done for 1D data of error matrix and then 1D data turned to 2D data by replacing m/n with $r$, $r=\sqrt{(m\Delta)^2 + (n\Delta)^2 }$}. After obtaining the best fit, the Gaussian phase autocorrelation matrix $B_{gauss}(m,n)$ can be obtained with the help of Eqn. \ref{eqn:sf_ac} by equating the zeroth component of the autocorrelation matrix to zero. The autocorrelation matrix after doing the Gaussian compensation can be given as

\begin{equation}
        B_{tot}(m,n) = B_{\phi}(m,n) + B_{gauss}(m,n).
\end{equation}

\section{Phase screen simulation using $B_{\lowercase{tot}}(\lowercase{m},\lowercase{n})$ matrix}\label{sec:phase_screen}
\begin{figure}[h!]
    \centerline{\includegraphics[width=0.99\textwidth]{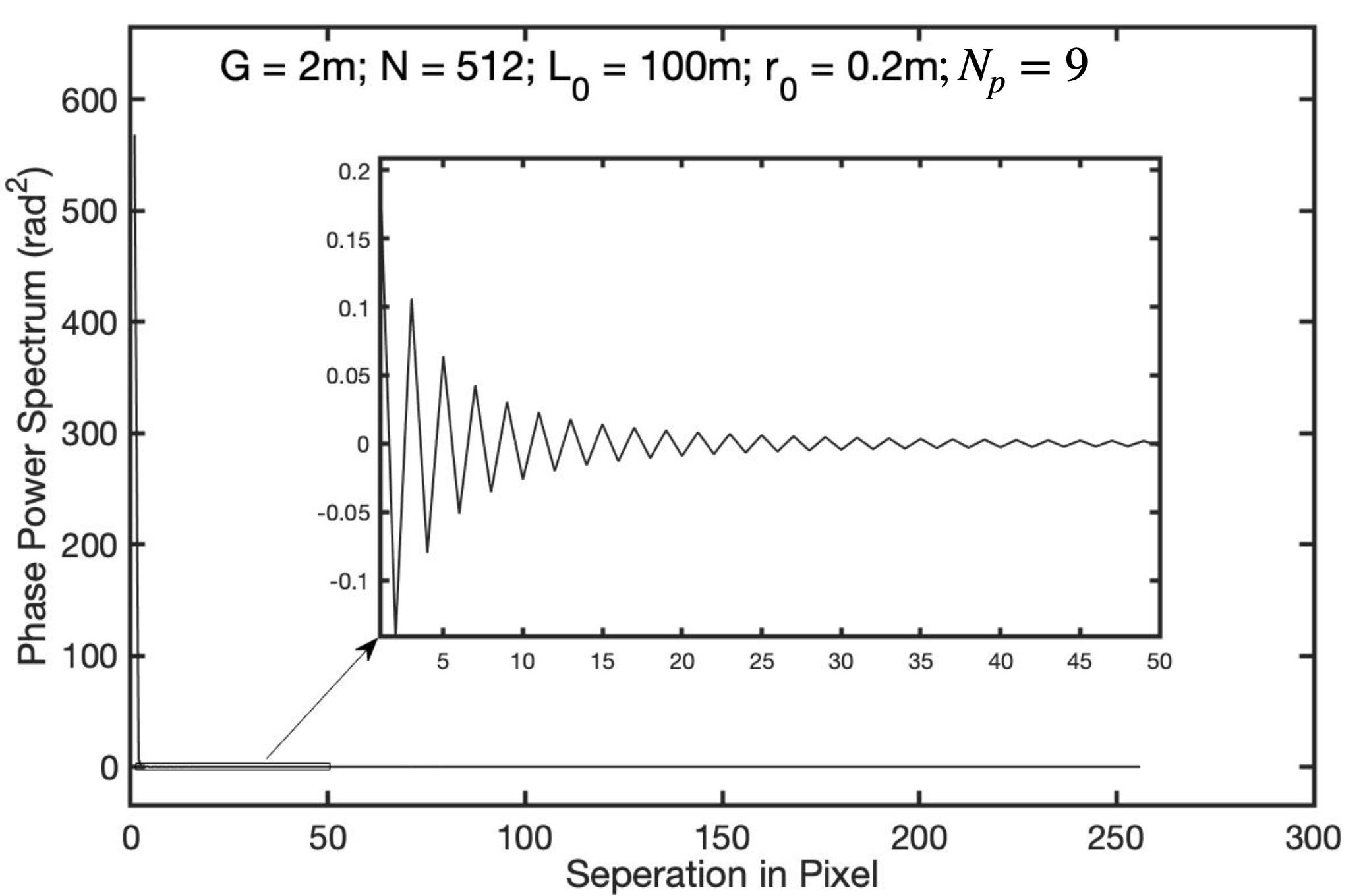}}
    \caption{\label{fig:power_spectrum}Negative Power spectrum Values for smaller $G/L_0$ ratio }
\end{figure}

Obtaining Power spectrum from $B_{tot}(m,n)$ will lead to negative terms in the power spectrum \cite{Xiang14}. Directly putting those frequency terms equal to zero leads to a loss in the energy spectrum. From Fig~\ref{fig:power_spectrum}, for the case of  $G/L_0<1$, we see that there is a large number of negative frequency components over a small separation. Hence $B_{tot}$(m,n) matrix needs to be preprocessed to eliminate most of these negative values in the power spectrum. For this we extract the Piston and Tip/Tilt components from the phase autocorrelation matrix $B_{tot}$. The Tip/Tilt component from phase autocorrelation matrix is given as

\begin{equation}
            B_{tilt}(r) = B_{tilt}(0) - r^2\sigma_{tilt}^2/2 
        \end{equation}
where $\sigma^2_{tilt}$ is the variance of the random tilt angle in the x or y directions and given as follows\cite{Xiang14}:

\begin{equation}
            \sigma^2_{tilt} = \frac{B_{tot}(G/2+\Delta)-B_{tot}(G/2)}{\Delta(G-\Delta)/2}   
\end{equation}
The remaining phase autocorrelation matrix of higher order terms is given as follows:

\begin{equation}
            B_{high}(r) = B_{tot}(r) - B_{tilt}(r)
\end{equation}
Standard fast fourier transform relation is then used in order to obtain the power spectrum from $B_{high}(r)$ and $B_{tilt}(r)$ as $f_{high}^2$ and $f_{tilt}^2$ respectively. Setting the negative values in $f_{high}^2$ and $f_{tilt}^2$ to zero leads a residual error that can further reduced by using a smoothening Gaussian operator as done by Xiang with updated parameters A=3.1 and width W = G/1.5.We have obtained these numbers after performing error analysis for different values of A and W over N = 128 sampling size phase screen and find for minimum relative error in structure function calculation.  To obtain the phase screen from the power spectrum, the following relation is used\cite{Xiang14}:
\begin{equation}
            \begin{aligned} 
            \phi(m \Delta, n \Delta)= \sum_{m^{\prime}=-N / 2}^{N / 2-1} \sum_{n^{\prime}=-N / 2}^{N / 2-1}\left[R_{a}\left(m^{\prime}, n^{\prime}\right)+\mathrm{i} R_{b}\left(m^{\prime}, n^{\prime}\right)\right] f\left(m^{\prime} \Delta^{\prime}, n^{\prime} \Delta^{\prime}\right) \exp \left[\mathrm{i} 2 \pi\left(m^{\prime} m+n^{\prime} n\right) / N\right] 
            \end{aligned}
\end{equation}
Where $R_a(m^{'},n^{'})$ and $R_b(m^{'},n^{'})$ are zero-mean and unity-variance gaussian random number generators. $f$ being $f_{high}$ and $f_{tilt}$, for the higher order terms and tip/tilt term respectively.

\section{Validation via phase structure function calculation}\label{sec:val}
\begin{figure}[h!]
    \centerline{\includegraphics[width=0.99\textwidth]{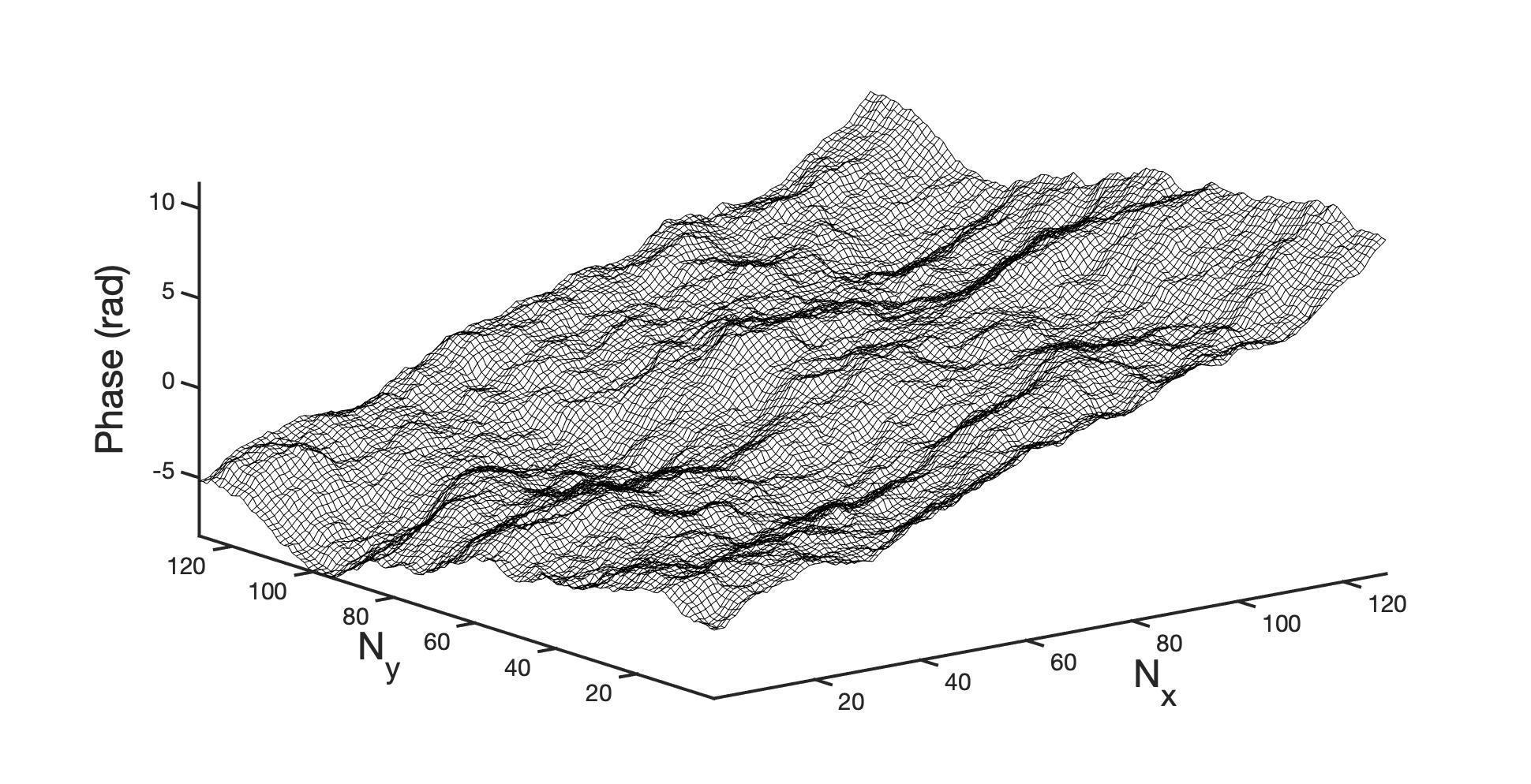}}
    \caption{\label{fig:phasescreen}Phase Screen for case $G=1 m$, $L_0 =100 m$ , $r_0 = 0.2 m$,$N_p = 3$, $N_x=N_y=128$ }
\end{figure}

\begin{figure}[h!]
    \centerline{\includegraphics[width=0.75\textwidth]{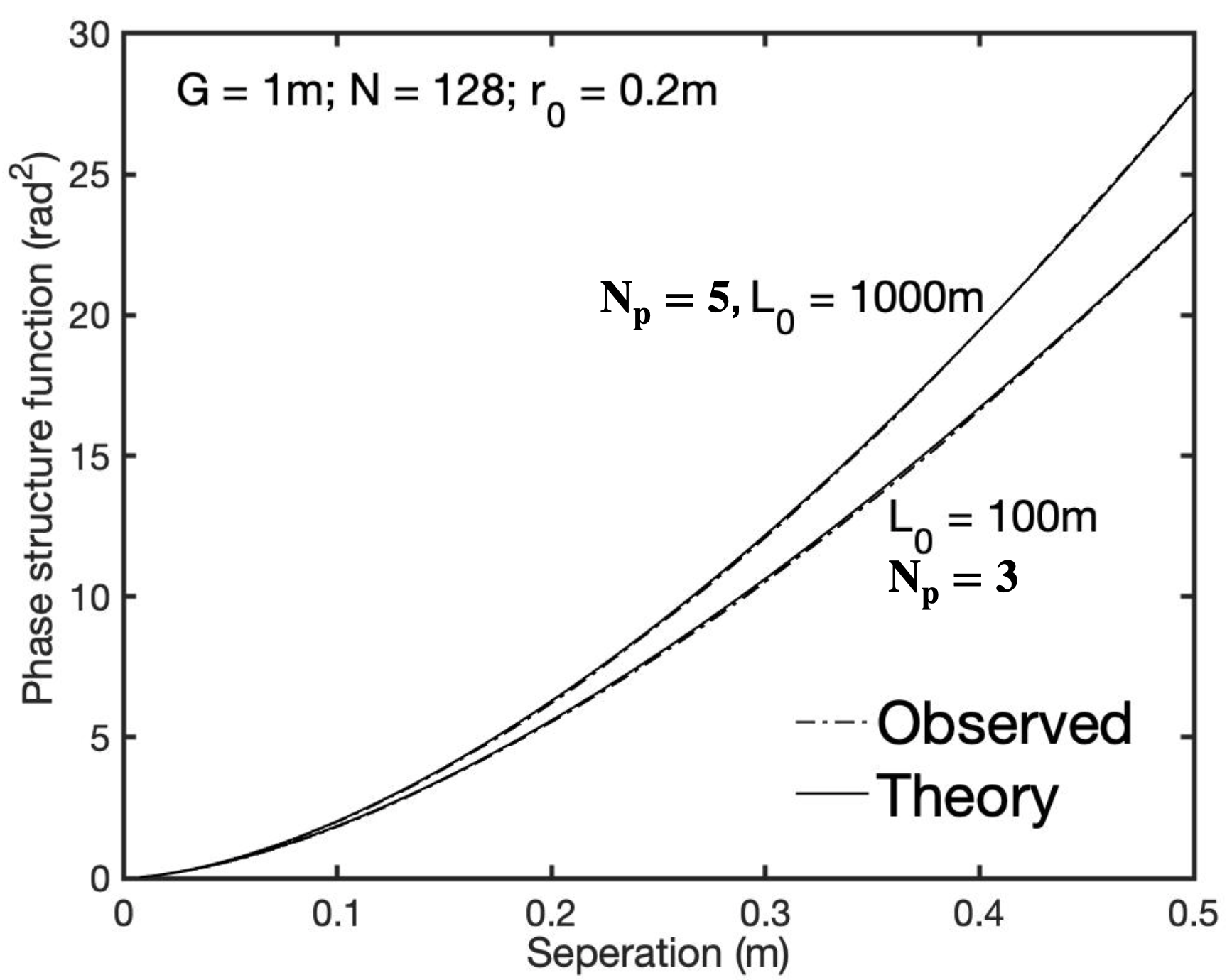}}
    \caption{\label{fig:psf}Structure function Calculation}
\end{figure}

\begin{figure}[h!]
    \centerline{\includegraphics[width=0.75\textwidth]{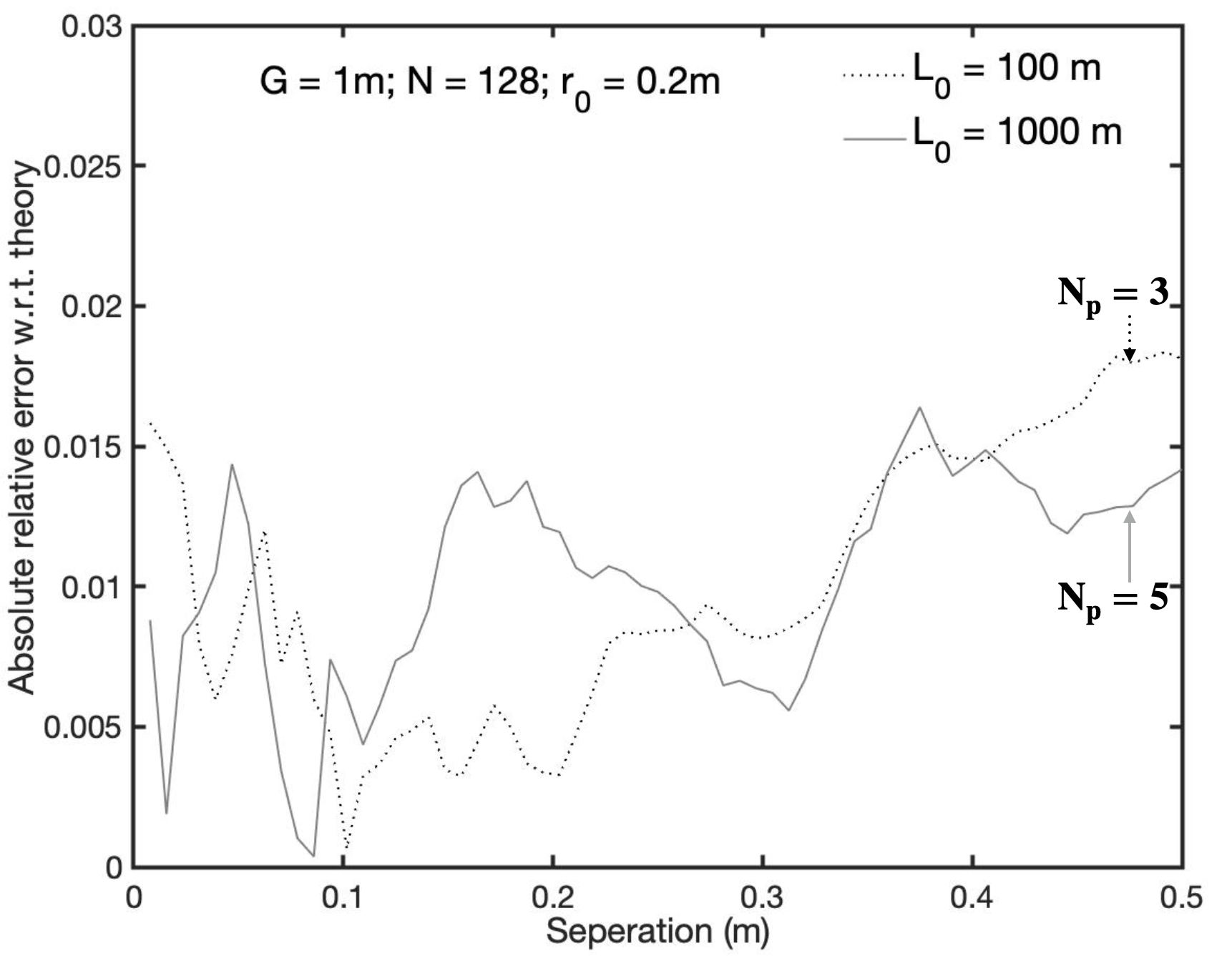}}
    \caption{\label{fig:error}Relative error w.r.t. Theory}
\end{figure}

Consider the parameters: G = 1m , $L_0$ = 100/1000 m , N = 128 , $r_0$= 0.2 m , A = 3.1 , W = G/1.5, $N_{p}=5/3$ and results has been averaged over 50K frames. Fig~\ref{fig:phasescreen} shows the corresponding phase screen plot for $L_0 = 100  m $. The Relative error in phase structure function calculation has been calculated as follows:
\begin{equation}
       err(r)= \frac{D_{\phi}^{obs}(r)-D_{theory}(r)}{D_{theory}(r)}
    \end{equation}
Here, $D_{\phi}^{obs}(r)$ is the Observed phase structure function from simulated phase screen.
The $max[err(r)]$ turns out to be $<1.8\%$ for $r<G/2$ as shown in the Fig~\ref{fig:error} and the corresponding phase structure function plot is shown in Fig~\ref{fig:psf}. 

\section{Result Analysis}\label{sec:results}
Our errors little shoot up from 1\% because we have not set phase autocorrelation matrix to zero for $r>G/2$. The reason is the following: After extracting piston and tilt from $B_{tot}(r)$(which has inherent residual error close to 0.5\% from curve fitting),$B_{high}(r)$ does not fall to zero immediately. Due to which there is a sudden jump. Further improvement that can be performed is to find perfect smoothening operators like Gaussian, which we can multiply with $B_{high}$ so that it falls to zero progressively, not sharply. This can provide further significant improvement in the final result. 

\section{Conclusion}\label{sec:conclusion}
In this paper, we put forward a new method to compensate for the residual error
in the Low and/or High-frequency region of FFT simulated phase screens after compensating with the modified subharmonic method. This method provides very
accurate phase screen structure for even $G/L_0$ ratios as small as 1/1000. No Patch Normalization factor is needed, no need to calculate subharmonic weight coefficient \cite{Lane92} and weights to compensate for high-frequency components, as done by Sedmak. Currently we have demonstrated this technique for only  circular screens. We have used MATLAB R2018b on Macbook pro 2018 with 8Gb Ram to calculate coefficients of the Gaussian matrix, which works pretty fast. Finally, the accuracy of this method from low-frequency to the high-frequency range is better than 1.8\% for $G/L_0$ as low as 1/1000.

\acknowledgments 
 We would like to thank Sedmak to support us over private communication and provide in-depth knowledge of the atmospheric power spectrum. We also thank Xiang for sharing his MATLAB code which calculates the phase structure function quickly for a large number of phase screens.

\bibliography{report}   
\bibliographystyle{spiebib} 

\end{document}